\documentclass[useAMS,usenatbib,onecolumn]{mn2e}
\usepackage{amssymb}
\usepackage{graphicx}
\usepackage{epsfig}
\usepackage{graphics}
\usepackage{bm}
\usepackage{mathptmx}
\usepackage[dvips]{color}
\expandafter\ifx\csname package@font\endcsname\relax\else
\expandafter\expandafter
\expandafter\usepackage
\expandafter\expandafter
\expandafter{\csname package@font\endcsname}%
\fi
\def\Quadrat#1#2{{\vcenter{\hrule height #2
\hbox{\vrule width #2 height #1 \kern#1
\vrule width #2}\hrule height #2}}}
\def\Box{\mathop{\kern 1pt\hbox{$\Quadrat{8pt}{0.4pt}$} \kern 1pt}}
\newcommand{\prd}{Phys. Rev. D}

\newcommand{\solphys}{Sol. Phys.}
\newcommand{\aj}{Astron. J. (USA)}
\title[Post-Newtonian limitations on measurement of the PPN
parameters]{Post-Newtonian limitations on measurement of the PPN parameters
caused by motion of gravitating bodies}
\author[S. M. Kopeikin]{S. M. Kopeikin\thanks{E-mail: kopeikins@missouri.edu}\\\\
Department of Physics \& Astronomy, University of
Missouri-Columbia, Columbia, MO 65211, USA}
\begin{document}

\date{Accepted 2009 July 8.  Received 2009 July 8; in original form 2009 June 9}

\pagerange{\pageref{firstpage}--\pageref{lastpage}} \pubyear{2009}

\maketitle

\label{firstpage}

\begin{abstract}
We derive explicit Lorentz-invariant solution of the Einstein and null
geodesic equations for data processing of the time delay and ranging
experiments in gravitational field of moving gravitating bodies of the solar
system - the Sun and major planets. We discuss general-relativistic
interpretation of these experiments and the limitations imposed by motion of
the massive bodies on measurement of the parameters $\gamma_{PPN}$, $\beta_{PPN}$ and
$\delta_{PPN}$ of the parameterized post-Newtonian formalism.
\end{abstract}
\section{Introduction}
Theoretical speculations beyond the Standard Model suggest that gravity must
be naturally accompanied by a partner - one or more scalar fields, which
contribute to the hybrid metric of space-time through a system of equations
of a scalar-tensor gravity theory \citep{sttg}. Such scalar partners
generically arise in all extra-dimensional theories, and notably in
string theory. Scalar fields play also an important role in modern
cosmological scenarios with the inflationary stage \citep{mukhanov}.
Therefore, unambiguous experimental verification of existence of the scalar
fields is among primary goals of gravitational physics.

Phenomenological presence of the scalar field in the metric tensor is
parameterized by three parameters -- $\gamma_{PPN}$, $\beta_{PPN}$ and $\delta_{PPN}$ -- of
the parameterized post-Newtonian (PPN) formalism. These parameters enter the
metric tensor of a {\it static} and {\it spherically-symmetric} gravitating
body in the following form \citep{dametal,brum,w-book,will-livrev}
\begin{eqnarray}
\label{aur1}
g_{00}&=&-1+\frac{2GM}{c^2 R}-2(1+\bar\beta_{PPN})\left(\frac{GM}{c^2
R}\right)^2+O\left(c^{-6}\right)\;,\\\label{aq2}
g_{ij}&=&\delta_{ij}\left[2(1+\bar\gamma_{PPN})\frac{GM}{c^2
R}+\frac{3}{2}(1+\bar\delta_{PPN})\left(\frac{GM}{c^2
R}\right)^2\right]+O\left(c^{-4}\right)\;,
\end{eqnarray}
where we have used the isotropic coordinates $X^\alpha=(cT,{\bm X})$,
$R=|{\bm X}|$, and denoted deviation from general relativity with the
comparative PPN parameters $\bar\gamma_{PPN}\equiv\gamma_{PPN}-1$,
$\bar\beta_{PPN}\equiv\beta_{PPN}-1$, $\bar\delta_{PPN}=\delta_{PPN}-1$. Parameter $\bar\delta_{PPN}$ generalizes the standard
PPN formalism \citep{w-book} to the second post-Newtonian approximation
\citep{brum}. One notices that $\delta_{PPN}$ is
actually related to $\beta_{PPN}$ and $\gamma_{PPN}$ in a generic scalar-tensor theory of gravity \citep{dametal}. In particular, this theory predicts that $\beta_{PPN}$ cancels
in the combination $-\beta_{PPN} + 3/4 \delta_{PPN}$ entering equation. (4) of the present paper, which
should depend, theoretically, only on $\gamma_{PPN}$ and its square. Thus, high-precision missions will have a very clean access to $\gamma_{PPN}$. However, we prefer to keep combination $-\beta_{PPN} + 3/4 \delta_{PPN}$ explicitly in our equations in order to separate parametrization of the second post-Newtonian effects associated with $\delta_{PPN}$ from the linearized Shapiro time delay, which is parameterized by $\gamma_{PPN}$ alone. Moreover, parameter $\delta_{PPN}$ is independent from $\beta_{PPN}$ and $\gamma_{PPN}$ in vector-tensor theories of gravity \citep{xie}. In general relativity, $\bar\beta_{PPN}=\bar\gamma_{PPN}=\bar\delta_{PPN}=0$.

The best experimental bound on $\bar\gamma_{PPN}=(2.1 \pm 2.3)\times 10^{-5}$ has
been obtained (under a certain implicit assumption \citep{kpsv}) in the Cassini experiment \citep{b0}. Limits on the
parameter $\bar\beta_{PPN}$ depend on the precision in measuring $\bar\gamma_{PPN}$, and
are derived from a linear combination $2\bar\gamma_{PPN}-\bar\beta_{PPN}<3\times
10^{-3}$ by observing the Mercury's perihelion shift, and from
$4\bar\beta_{PPN}-\bar\gamma_{PPN} = (4.5\pm 4.5)\times 10^{-4}$ imposed by the lunar
laser ranging \citep{wtb}. Parameter $\bar\delta_{PPN}$ has not yet been measured.

The most precise measurement of $\bar\gamma_{PPN}$ and $\bar\delta_{PPN}$ can be
achieved in near-future gravitational experiments with light propagating in
the field of the Sun or a major planet.
Post-post-Newtonian equation of the relativistic time delay in a static
gravitational field is obtained from the metric (\ref{aur1}), (\ref{aq2}).
It was derived by a number of authors
\citep{1982PhRvD..26.2549R,1983PhRvD..28.3007R,brum,2008arXiv0803.0277T} and
reads (in the isotropic coordinates) as follows
\begin{equation}
\label{aq3}
T_2-T_1=\frac{R}{c}+\Delta T+O(G^3)\;,
\end{equation}
where $T_1$ and $T_2$ are coordinate times of emission and observation of
photon, $R=|{\bm X}_2-{\bm X}_1|$ is the coordinate distance between the
point of emission, ${\bm X}_1$, and observation, ${\bm X}_2$, of the photon,
and
\begin{eqnarray}
\label{aq4}
\Delta
T&=&(2+\bar\gamma_{PPN})\frac{GM}{c^3}\ln\left(\frac{R_1+R_2+R}{R_1+R_2-R}\right)
\\\nonumber
&+&\frac{G^2M^2}{c^5}\frac{R}{R_1R_2}\left[\left(\frac{15}{4}+2\bar\gamma_{PPN}-\bar
\beta_{PPN}+\frac{3}{4}\bar\delta_{PPN}\right)\frac{\arccos({\bm N}_1\cdot{\bm
N}_2)}{|{\bm N}_1\times{\bm N}_2|}-\frac{(2+\bar\gamma_{PPN})^2}{1+{\bm
N}_1\cdot{\bm N}_2}\right]
\end{eqnarray}
is the extra time delay caused by the gravitational field, ${\bm N}_1={\bm
X}_1/R_1$ and ${\bm N}_2={\bm X}_2/R_2$ are the unit vectors directed
outward of the gravitating body, $R_1=|{\bm X}_1|$, $R_2=|{\bm X}_2|$ are
radial distances to the points of emission and observation respectively.

The Sun and planets are not at rest in the solar system because they are
moving with respect to the barycenter of the solar system as well as with
respect to observer. Motion of the light-ray deflecting body (the Sun, a
major planet) affects propagation of light bringing the post-Newtonian
corrections of the order of $(GM/c^3)(v/c)$, $(GM/c^3)(v/c)^2$, etc. to
equation (\ref{aq4}), where $v$ is a characteristic speed of the massive
body with respect to a reference frame used for data processing, which can
be chosen as either the barycentric frame of the solar system or the
geocentric frame of observer. These motion-induced post-Newtonian
corrections to the static time delay $\Delta T$ correlate with the PPN
parameters making their observed numerical value biased. Therefore, it is
important to disentangle the genuine effects associated with the presence of
the scalar field from the special-relativistic effects in equation
(\ref{aq4}) imparted by the motion of the bodies.

This problem has not been addressed until recently because the accuracy of
astronomical observations was not high enough. However, VLBI measurement of
the null-cone gravity-retardation effect \citep{kop2001,fk-apj,kop-cqg,fom09} and
frequency-shift measurement of $\gamma_{PPN}$ in the Cassini experiment
\citep{a04,b0} made it evident that modern technology has achieved the level
at which relativistic effects caused by the dependence of the gravitational
field on time can be no longer ignored. Future gravitational light-ray
deflection experiments \citep{km}, radio ranging  BepiColombo experiment
\citep{2002PhRvD..66h2001M}, laser ranging experiments ASTROD
\citep{2007NuPhS.166..153N} and LATOR \citep{2004CQGra..21.2773T} will
definitely reach the precision in measuring
$\bar\gamma_{PPN}$, $\bar\beta_{PPN}$ and $\bar\delta_{PPN}$ that is comparable with the
post-Newtonian corrections to the static time delay and to the deflection
angle caused by the motion of the massive bodies in the solar system  \citep{2006CQGra..23..309P}.
Therefore, it is worthwhile to undertake a scrutiny theoretical study of the
time-dependent relativistic corrections to the static Shapiro time delay.

In this paper we focus on deriving two apparently different forms of the Lorentz invariant solution of the
light ray equations (see equations (31) and (43)) in the linearized (with respect to the universal
gravitational constant G) approximation of general relativity by making use
of the technique of the Li\'enard-Wiechert potentials \citep{ks} and algebraic transformations of the retarded quantities. In particular, equation (43) of the present paper significantly generalizes the result of \citet{bai} for the gravitational time delay. We expand
this retarded-time solution in the post-Newtonian series in three various ways (see equations (63), (68) and (86) below) and analyze the impact of the
velocity-dependent corrections on measuring values of the PPN parameters in
the gravitational time-delay experiments. Section VIII discusses a correspondence between the Lorentz symmetry group for gravity and light as revealed by the time delay experiments. Section IX gives a justification that the ODP code of NASA must be revamped for doing adequate processing of high-precise data in ranging gravitational experiments.

\section{Notations}\label{intro}

In what follows the Greek indices $\alpha, \beta,...$ run from 0 to 3,
the Roman indices $i,j,...$ run from 1 to 3, repeated Greek indices
mean Einstein's summation from 0 to 3, and bold letters ${\bm
a}=(a^1,a^2,a^3), {\bm b}=(b^1,b^2,b^3),$ etc. denote spatial
(3-dimensional) vectors. A dot between two spatial vectors, for example
${\bm a}\cdot{\bm b}=a^1b^1+a^2b^2+a^3b^3$, means the Euclidean dot
product, and the cross between two vectors, for example ${\bm
a}\times{\bm b}$, means the Euclidean cross product. We also use a
shorthand notation for partial derivatives $\partial_\alpha
=\partial/\partial x^\alpha$. Greek indices are raised and lowered with
full metric $g_{\alpha\beta}$. The Minkowski (flat) space-time metric
$\eta_{\alpha\beta}={\rm diag}(-1,+1,+1,+1)$. This metric is used to rise
and lower indices of the unperturbed wave vector $k^\alpha$ of light, and
the gravitational  perturbation $h_{\alpha\beta}$.

\section{The Li\'enard-Wiechert Gravitational Potentials}
We introduce the post-Minkowskian decomposition of
the metric tensor
\begin{equation}
\label{y}
g_{\alpha\beta}=\eta_{\alpha\beta}+h_{\alpha\beta}\;,
\end{equation}
where $h_{\alpha\beta}$ is the post-Minkowskian perturbation of the
Minkowski metric tensor $\eta_{\alpha\beta}$. We impose the harmonic
gauge condition \citep{mtw} on the metric tensor
\begin{equation}
\label{hgc}
\partial_\alpha h^{\alpha\beta}-\frac{1}{2}\partial^\beta
h^{\lambda}_{\;\,\lambda}=0\;.
\end{equation}
In arbitrary harmonic coordinates $x^\alpha=(ct,
{\bm x})$, and in the first post-Minkowskian approximation the Einstein
equations read
\begin{eqnarray}
\label{gfe}
\left(-\frac{1}{c^2}\frac{\partial^2}{\partial t^2}
+\nabla^2\right)h^{\mu\nu}&=&-\frac{16\pi G
}{c^4}\left(T^{\mu\nu}-\frac{1}{2}\eta^{\mu\nu}T^\lambda_{\;\,\lambda}\right
)\;.
\end{eqnarray}
where $T^{\mu\nu}$ is the stress-energy
tensor of a light-ray deflecting body. In linearized approximation this
tensor is given by the following equation
\begin{eqnarray}
\label{hh}
T^{\mu\nu}(t, {\bm x})=Mu^\mu
u^\nu \sqrt{1-\beta^2}\delta^{(3)}\bigl({\bm x}-{\bm z}(t)\bigr)\;,
\end{eqnarray}
where $M$ is the (constant) rest mass of the body,
${\bm z}(t)$ is time-dependent spatial coordinate of the body,
${\bm\beta}= c^{-1}d{\bm z}/dt$ is velocity of the body normalized to the
fundamental speed $c$,
\begin{equation}
u^0=\left(1-\beta^2\right)^{-1/2}\qquad,\qquad
u^i=\beta^i\left(1-\beta^2\right)^{-1/2}\;,
\end{equation} is the four-velocity of the
body normalized such that $u_\alpha u^\alpha=-1$, and $\delta^{(3)}({\bm
x})$ is the 3-dimensional
Dirac's delta-function.  We have neglected $\sqrt{-g}$ in equation
(\ref{hh}) because in the linearized approximation $\sqrt{-g}=1+O(G)$,
and the quadratic terms proportional to $G^2$ are
irrelevant in $T^{\mu\nu}$ since they will give time-dependent terms of the
second post-Minkowskian order of magnitude, which are currently negligible for
measurement in the solar system. For the same reason, we do not use the metric derived by \citet{blanch} as it goes beyond  the approximation used in the present paper.
We have also used a standard notation ${\bm\beta}$ for the dimensionless velocity of the body. This notation should not be confused with the PPN parameter $\beta_{PPN}$.

Because the Einstein equations (\ref{gfe}) are linear, we can consider
their solution as a linear superposition of the solutions for each
body. It allows us to focus on the relativistic effects caused by one
body (the Sun, planet) only.
Solving Einstein's equations (\ref{gfe}) by making use of the retarded
Li\'enard-Wiechert tensor potentials \citep{bd}, one obtains the
post-Minkowski metric
tensor perturbation \citep{bd,ks}
\begin{equation}\label{1}
h^{\mu\nu}(t,{\bm x})=\frac{4GM}{c^2}\frac{u^\mu
u^\nu+\frac12\eta^{\mu\nu}}{\rho_R}\;,
\end{equation}
where
\begin{eqnarray}
\label{1q}
\rho_R&=& -u_\alpha \rho^\alpha\;,\\\label{1w}
\rho^\alpha&=&x^\alpha-z^\alpha(s)\;.
  \end{eqnarray}
  In equation (\ref{1}) all time-dependent quantities are taken at retarded
time $s$ defined by the null cone equation (\ref{grav}) given below,
$u^\alpha\equiv u^\alpha(s)=c^{-1}dz^\alpha(s)/ds$ is its four-velocity,
with $s$ being a retarded
time (see below), ${\bm\beta}(s)=c^{-1}d{\bm z}(s)/ds$ is
body's coordinate velocity normalized to the fundamental speed $c$. Notice
that the metric
tensor perturbation (\ref{1}) is valid for accelerated motion of the
gravitating body as well, and is not restricted by the approximation of a
body moving on
a straight line (see \citet{bd} for more detail). In other words, the
four-velocity $u^\alpha$ in equation (\ref{1}) is not a constant, taken at
one, particular event on the world line of the body.

Because we solved the Einstein equations
(\ref{gfe}) in terms of the retarded Li\'enard-Wiechert potentials, the
distance $\rho^\alpha=x^\alpha-z^\alpha(s)$, the
body's worldline $z^\alpha(s)=(cs, {\bm z}(s))$, and the
four-velocity $u^\alpha(s)$ are all functions of the {\it retarded} time
$s$ \citep{bd}. The retarded time $s$ is found in the first post-Minkowski
approximation as a solution of the {\it null cone} equation
\begin{equation}
\label{grav}
\eta_{\mu\nu}\rho^\mu
\rho^\nu\equiv\eta_{\mu\nu}\Bigl(x^\mu-z^\mu(s)\Bigr)\Bigl(x^\nu-z^\nu(s)\Bigr)=0\;,
\end{equation}
that is
\begin{equation}
\label{1a}
s=t-\frac{1}{c}|{\bm x}-{\bm z}(s)|\;,
\end{equation}
where the constant $c$ in equation (\ref{1a}) denotes the fundamental speed in
the Minkowski space-time, which physical meaning in equation (\ref{1a}) is the speed of propagation of gravity as it originates from the gravity field equations (\ref{gfe}). It is important to notice that equation (\ref{1a})
is a complicated function of the retarded time $s=s(t,{\bm x})$, which has an analytic solution only in case of a uniform motion of the gravitating body along a straight line \citep{kop-cqg}. Geometrically, equation (\ref{1a}) connects the point of observation ${\bm x}$ and the retarded position
of the gravitating body ${\bm z}(s)$ by a null
characteristic of the linearized Einstein field equations (\ref{gfe}). Radio waves
(light) are also
propagating along a null characteristic connecting the observer and the radio emitter.
However, the null characteristic of
the linearized Einstein equations (\ref{1a}) is well separated on the space-time manifold (and in the sky)
from the
null characteristic associated with the propagation of the radio wave in any kind of
ranging and time-delay experiments. Hence,
they should not be confused in relativistic experiments involving light
propagation in the field of a moving gravitating body, which gravitational
field depends on time \citep{will-livrev,kffp}.

All components of the time-dependent gravitational field (the metric tensor perturbation $h_{\alpha\beta}$) of the solar
system bodies interact with radio (light) waves moving from a radio (light) source to the Earth,
and perturb each element of the phase of electromagnetic wave with the retardation given by equation (\ref{1a}).  The use of the
retarded Li\'enard-Wiechert gravitational potentials, rather than the
advanced
potentials, is consistent with the principle of causality \citep{kop-fom}, and the
observation of the orbital decay of the relativistic binary pulsar B1913+16
caused by the emission of gravitational
radiation, according to general relativity \citep{wt}.

\section{The electromagnetic phase}

Any ranging or time delay experiment measures the phase $\psi$
of an electromagnetic wave coming from a spacecraft or a radio (light) source
outside of the solar system.  The phase is a scalar function being invariant
with respect to coordinate transformations.
It is determined in the approximation of geometric optics from the
eikonal equation \citep{mtw,ll}
\begin{equation}
\label{eik}
g^{\mu\nu}\partial_\mu\psi\partial_\nu\psi=0\;,
\end{equation}
where $g^{\mu\nu}=\eta^{\mu\nu}-h^{\mu\nu}$.
The eikonal equation (\ref{eik}) is a
direct consequence of Maxwell's equations \citep{mtw,km} and its
solution describes localization of the front of an electromagnetic wave propagating
on a curved space-time manifold, which geometric properties are defined by the metric tensor
(\ref{y}), (\ref{1}) that is a solution of the Einstein equations. We
emphasize that the electromagnetic wave in
equation (\ref{eik}) has no back-action on the
properties of the metric tensor $g_{\mu\nu}$, and does
not change the curvature of the space-time caused by the presence of the
gravitating body. Thus, experimental studying of the propagation of
electromagnetic wave allows us to measure the important properties of the
background gravitational field and space-time manifold.

Let us introduce a co-vector of the electromagnetic wave,
$K_\alpha=\partial_\alpha\psi$. Let
$\lambda$ be an affine parameter along a light ray being orthogonal to the
electromagnetic wave front $\psi$. Vector
$K^\alpha=dx^\alpha/d\lambda=g^{\alpha\beta}\partial_\beta\psi$ is
tangent to the light ray. Equation (\ref{eik}) expresses a simple fact
that vector $K^\alpha$ is null, that is $g_{\mu\nu}K^\mu
K^\nu=0$. Thus, the light rays are null geodesics \citep{ll} defined by
equation
\begin{equation}
\label{geo}
\frac{dK_\alpha}{d\lambda}=\frac{1}{2}\partial_\alpha g_{\mu\nu}K^\mu
K^\nu\;.
\end{equation}
The eikonal equation (\ref{eik}) and light-ray equation (\ref{geo}) have
equivalent physical content in general relativity since equation (\ref{eik})
is a first integral of equation (\ref{geo}).

Regarding propagation of electromagnetic wave, it is more straightforward to
find solution of equation (\ref{eik}). To this end, we expand the eikonal
$\psi$ in the post-Minkowskian series with respect to the universal
gravitational constant $G$ assuming that the unperturbed solution of
equation (\ref{eik}) is a plane electromagnetic wave (that is, the parallax
of the radio source is neglected). The expansion reads
\begin{equation}
\label{ei4}
\psi=\psi_0+\frac{\nu}{c}\left[ k_\alpha x^\alpha+\varphi(x^\alpha)\right]+O(G^2)\;,
\end{equation}
where $\psi_0$ is a constant of integration,
$k^\alpha=(1,{\bm k})$ is a constant null vector directed along the
trajectory of propagation of the
unperturbed electromagnetic wave such that $\eta_{\mu\nu}k^\mu k^\nu=0$,
$\nu$ is  the constant frequency of the unperturbed electromagnetic wave,
and $\varphi$ is the first post-Minkowskian perturbation of the eikonal,
which is Lorentz-invariant. Substituting expansions (\ref{y}), (\ref{ei4})
to equation (\ref{eik}), and leaving only terms of order $G$, one obtains an
ordinary differential equation for the post-Minkowskian perturbation of the
eikonal,
\begin{equation}
\label{asd}
\frac{d\varphi}{d\lambda}=\frac{1}{2}h^{\alpha\beta}k_\alpha
k_\beta=\frac{2GM}{c^2}\frac{(u_\alpha k^\alpha)^2}{\rho_R}\;,
\end{equation}
which can be also obtained as a first integral of the null geodesic equation
(\ref{geo}).
Equation (\ref{asd}) can be readily integrated if one employs an exact
relationship
\begin{equation}
\label{br}
\frac{d\lambda}{\rho_R}=-\frac{ds}{k_\alpha \rho^\alpha}=\frac{1}{k_\alpha
u^\alpha}\,d\Bigl[\ln\left(-k_\alpha \rho^\alpha\right)\Bigr]\;,
\end{equation}
which makes the integration straightforward. Indeed,
if the body's acceleration is neglected, a plane-wave solution of equation
(\ref{asd}) is
\begin{equation}
\label{3}
\varphi(x^\alpha)=\frac{2GM\nu}{c^3}\left(k_\alpha
u^\alpha\right)\ln\left(-k_\alpha \rho^\alpha\right)\;,
\end{equation}
where all quantities in the right side are taken at the retarded instant of
time $s$ in compliance with the null cone equation (\ref{1a}). One notices that the time $t_*$ of the closest approach of the light ray to the moving body does not play any role in calculation of the gravitational perturbation of the electromagnetic phase. The time $t_*$ is a good approximation of the retarded time $s$ \citep{kop2001}, and can be used in practical calculations of light propagation in the gravitational field of moving bodies \citep{klikop,grem}. However, it does not properly reflect the Lorentz-invariant nature of the gravitational time delay and makes its post-Newtonian expansion looking more entangled and complicated. Further discussion of this issue is given in section \ref{icoup}.

One can easily check that equation (\ref{3}) is a particular solution
of equation (\ref{eik}). Indeed, observing
that
\begin{equation}
\label{qa}
\partial_\alpha \rho^\mu=\delta^\mu_\alpha-u^\mu\partial_\alpha s\;,
\end{equation}
one obtains from the null cone equation (\ref{grav})
\begin{equation}
\label{po}
\partial_\alpha s=-\frac{\rho_\alpha}{\rho_R}\;.
\end{equation}
Differentiation of equation (\ref{3}) using equations (\ref{qa}) and
(\ref{po}) shows that equation (\ref{eik}) is satisfied.

Equation (\ref{3}) for the electromagnetic phase is clearly
Lorentz-invariant and valid in an arbitrary coordinate system.  It tells us
that a
massive body  (the Sun, planet) interacts with the electromagnetic wave by
means of its gravitational field, which originates at the retarded position
${\bm z}(s)$ of the body and propagates on the hypersurface of null cone
(\ref{1a}). The gravitational field
perturbs the phase front of the electromagnetic wave at the field point
$x^\alpha$ regardless of the
direction of motion of the incoming photon or the magnitude of its
impact parameter with respect to the body. This consideration indicates a
remarkable experimental opportunity to observe the retardation effect of the
gravitational field by measuring the shape of the ranging
(Shapiro) time delay and comparing it with the JPL ephemeris position of the
body  \citep{standish} obtained independently from direct radio/optical
observations of the body, conducted in preceding epochs. This idea was
executed in VLBI experiment with Jupiter \citep{kop2001,fk-apj}. Next section
explains the null-cone relationship between the characteristics of the
Maxwell and Einstein equations.

\section{The Ranging Time Delay}

The Lorentz-invariant, general-relativistic time delay equation,
generalizing the
static Shapiro delay \citep{shap}, can be obtained directly from
equation (\ref{3}). We consider a ranging time-delay experiment in which an
electromagnetic wave (a photon) is emitted at the event with 4-dimensional
coordinates $x^\alpha_1=(ct_1,{\bm x}_1)$, passes near the moving
gravitating body, and is received by observer at the event with coordinates
$x^\alpha_2=(ct_2,{\bm x}_2)$. In the most general case, the emitter and
observer can move, which means that coordinates ${\bm x}_1$ and ${\bm x}_2$
must be understood as functions depending on time $t_1$ and $t_2$
respectively, that is ${\bm x}_1={\bm x}(t_1)$ and ${\bm x}_2={\bm x}(t_2)$,
where ${\bm x}(t)$ is a spatial coordinate of the photon taken at time $t$. The gravitating body is also
moving during the time of propagation of the electromagnetic wave from the emitter to the observer. In the
approximation of a uniform and rectilinear motion, which is sufficient for
our purpose, spatial coordinate of the body is given by a straight line
\begin{equation}
\label{yuw}
{\bm z}(t)={\bm z}_0+{\bm v}t\;,
\end{equation}
where ${\bm z}_0$ is position of the body taken time $t=0$. One notices that
the spatial coordinate of the body entering the Li\'enard-Wiechert solution of the gravity field equations depends on the retarded time $s$. It means that the time argument $t$ in equation (\ref{yuw}) must be replaced with the retarded time $s$ without changing the form of this equation. In other words,
\begin{equation}
\label{yuw1}
{\bm z}(s)={\bm z}_0+{\bm v}s\;,
\end{equation}
where the retarded time $s$ is given by the solution of the gravity null cone equation (\ref{1a}). In case of a rectilinear and uniform motion of the gravitating body
\begin{equation}
\label{solu}
s=t-\frac{{\bm R}\cdot{\bm\beta}+\sqrt{R^2-({\bm R}\cdot{\bm\beta})^2}}{c(1-\beta^2)}\;,
\end{equation}
and ${\bm R}={\bm x}-{\bm z}(t)$ with ${\bm z}(t)$ defined in equation (\ref{yuw}).

The unperturbed spatial components $(k^i)={\bm k}$ of the wave vector
$k^\alpha$ are expressed in terms of the coordinates of the emitting and
observing points
\begin{equation}
\label{utr}
{\bm k}=\frac{{\bm x}_2-{\bm x}_1}{|{\bm x}_2-{\bm x}_1|}\;.
\end{equation}
This vector is constant for a single passage of the electromagnetic wave
from the emitter to the observer. However, in case when the emitter and/or
observer are in motion, the direction of vector ${\bm k}$ will change as
time progresses. This remark is important for calculation of the Doppler
shift of frequency, where one has to take the time derivative of the vector
${\bm k}$ \citep{ks,kpsv}

The perturbed wave vector, $K^\alpha=dx^\alpha/d\lambda$, is obtained from
the eikonal equation (\ref{3}) by making use of identification
$K^\alpha=\partial\psi/\partial x^\alpha$, which is a consequence of the
Hamiltonian theory of light rays and can be used for further integration in
order to determine the trajectory of propagation of the electromagnetic wave
in the curved space-time. The explicit integration has been performed in
paper by \citet{kffp} and could be used for calculation of the ranging time
delay. However, in the present paper we shall rely upon a different method.

We note that the phase $\psi$ of the electromagnetic wave,
emitted at the point $x^\alpha_1=(ct_1,{\bm x}_1)$ and received at the
point $x^\alpha_2=(ct_2,{\bm x}_2)$, remains constant along the wave's path
\citep{mtw,ll,km}. Indeed, since $\lambda$ is an affine parameter along
the path, one has for the phase's derivative
\begin{equation}
\label{phase}
\frac{d\psi}{d\lambda}=\frac{\partial\psi}{\partial x^\alpha}\frac{
dx^\alpha}{d\lambda}=K_\alpha K^\alpha=0\;,
\end{equation}
which means that $\psi\left(x^\alpha(\lambda)\right)= ${\rm const.}, in
accordance with our assertion.  Equating two values of the phase $\psi$ at
the point of emission of the electromagnetic wave,
$x^\alpha_1$, and at the point of its receptions, $x^\alpha_2$, and
separating time from space coordinates,
one obtains from equations (\ref{ei4}), (\ref{3})
\begin{equation}
\label{tde0}
t_2-t_1=\frac{1}{c}{\bm k}\cdot\left({\bm x}_2-{\bm
x}_1\right)-\frac{2GM}{c^3}\left(k_\alpha
u^\alpha\right)\ln\left[\frac{k_\beta \rho^\beta_2}{k_\beta
\rho^\beta_2}\right]\;,
\end{equation}
where the retarded distances $\rho^\alpha_2=x^\alpha_2-z^\alpha(s_2)$,
$\rho^\alpha_1=x^\alpha_1-z^\alpha(s_1)$, and the retarded
times $s_2$, $s_1$ are defined by the null cone equations
\begin{eqnarray}
\label{we1}
s_2&=&t_2-\frac{1}{c}|{\bm x}_2-{\bm z}(s_2)|\;,\\\label{we2}
s_1&=&t_1-\frac{1}{c}|{\bm x}_1-{\bm z}(s_1)|\;,
\end{eqnarray}
which are inferred from equation (\ref{1a}).
Expanding all Lorentz-invariant scalar products, and replacing relationship
(\ref{utr}) in equation (\ref{tde0}) yields the ranging delay
\begin{eqnarray}
\label{tde}
t_2-t_1&=&\frac{1}{c}|{\bm x}_2-{\bm
x}_1|+\Delta t\;,\\\label{tdz}
\Delta t&=&-\frac{2GM}{c^3}\frac{1-{\bm
k}\cdot{\bm\beta}}{\sqrt{1-\beta^2}}\ln\left[\frac{\rho_2-{\bm
k}\cdot{\bm\rho}_2}{\rho_1-{\bm k}\cdot{\bm\rho}_1}\right]\;,
\end{eqnarray}
where the retarded, null-cone distances ${\bm\rho}_2={\bm x}_2-{\bm z}(s_2)$,
${\bm\rho}_1={\bm x}_1-{\bm z}(s_1)$, $\rho_2=|{\bm \rho}_2|$,
$\rho_1=|{\bm\rho}_1|$.

Lorentz-invariant expression for ranging
delay (\ref{tdz}) was derived first by
\citet{ks} by solving equations for light geodesics in the
gravitational field of moving bodies with the Li\'enard-Wiechert
gravitational potentials. Later on, \citet{kl} obtained
this expression by making use of the Lorentz transformation of the
Shapiro time delay (which is equivalent to a simultaneous transformation of
the solutions of both the Einstein and Maxwell equations) from a
static frame of the body to a moving frame of observer. Notice that in
general relativity equation (\ref{tde}) describes a
hypersurface of the null cone along which both electromagnetic and
gravitational field are propagating. Electromagnetic characteristic of the
null cone is given by the null vector ${\bm k}$ of the photon, while the
null characteristic of the gravity field enters the time delay equation
(\ref{tdz}) in the form of the retarded time $s$, which is the
time argument of the coordinate ${\bm z}$ of the moving body under consideration.

In the present paper we derive another useful form of the Lorentz-invariant
expression for the ranging delay, which can be directly compared with and generalizes the approximate
ranging delay formula currently used in the NASA Orbit Determination Program (ODP).
This derivation comes about from the following exact relationships
\begin{eqnarray}
\label{p1}
\rho_2-{\bm k}\cdot{\bm\rho}_2&=&\frac{|{\bm\rho}_1-{\bm z}(s_2)+{\bm
z}(s_1)|^2-(r-\rho_2)^2}{2r}\;,\\\label{p2}
\rho_1-{\bm k}\cdot{\bm\rho}_1&=&-\frac{|{\bm\rho}_2+{\bm z}(s_2)-{\bm
z}(s_1)|^2-(r+\rho_1)^2}{2r}\;,
\end{eqnarray}
where $r=|{\bm r}|$, ${\bm r}={\bm x}_2-{\bm x}_1$, so that
\begin{equation}
\label{p3}
r^\alpha=r k^\alpha=(r, {\bm r})\;,
\end{equation}
is a null vector in the flat space-time connecting coordinates of the point
of emission and reception of the electromagnetic wave:
$\eta_{\alpha\beta}r^\alpha r^\beta=0$. Because the gravitating body moves
uniformly with constant speed ${\bm v}$, its coordinate ${\bm z}(s)$ is not
constant and can be expanded as follows (see equation (\ref{yuw1}))
\begin{equation}
\label{p4}
{\bm z}(s_2)={\bm z}(s_1)+{\bm v}\left(s_2-s_1\right)\;,
\end{equation}
where the time interval $s_2-s_1$ can be expressed in terms of the null-cone distances by
making use of the retarded time equations (\ref{we1}), (\ref{we2}), and the
ranging equation (\ref{tde}). One has,
\begin{equation}
\label{p5}
s_2-s_1\equiv(s_2-t_2)+(t_2-t_1)+(t_1-s_1)=\frac1c\left(r+\rho_1-\rho_2\right)+O(c^{-3})\;.
\end{equation}
Plugging equation (\ref{p5}) to (\ref{p4}), and replacing it in equations
(\ref{p1}), (\ref{p2}) allows us to transform the ranging time delay
logarithm to the following form
\begin{equation}
\label{p6}
\ln\left[\frac{\rho_2-{\bm k}\cdot{\bm\rho}_2}{\rho_1-{\bm
k}\cdot{\bm\rho}_1}\right]=-\ln\left[\frac{\rho_2+\rho_1+r-2({\bm\rho_2}\cdot{\bm\beta})-\beta^2\left(r+\rho_1-\rho_2\right)}{\rho_2+\rho_1-r-2({\bm\rho_1}\cdot{\bm\beta})+\beta^2\left(r+\rho_1-\rho_2\right)}\right]\;.
\end{equation}
 Let us now make use of definition (\ref{1q}) of the Lorentz-invariant
distances
\begin{eqnarray}
\label{p7}
\rho_{2R}&=&-u_\alpha\rho^\alpha_2=\frac{\rho_2-{\bm\beta}\cdot{\bm\rho_2}}{
\sqrt{1-\beta^2}}\;,\\\label{p8}
\rho_{1R}&=&-u_\alpha\rho^\alpha_1=\frac{\rho_1-{\bm\beta}\cdot{\bm\rho_1}}{
\sqrt{1-\beta^2}}\;.
\end{eqnarray}
Tedious but straightforward calculations reveal that
\begin{eqnarray}
\label{p9}
\rho_2+\rho_1+r-2({\bm\rho_2}\cdot{\bm\beta})-\beta^2\left(r+\rho_1-\rho_2\right)=\sqrt{1-\beta^2}\left(\rho_{2R}+\rho_{1R}-rk_\alpha
u^\alpha\right)\;,\\\label{p10}
\rho_2+\rho_1-r-2({\bm\rho_1}\cdot{\bm\beta})+\beta^2\left(r+\rho_1-\rho_2\right)=\sqrt{1-\beta^2}\left(\rho_{2R}+\rho_{1R}+rk_\alpha u^\alpha\right)\;.
\end{eqnarray}

These equations taken along with equation (\ref{p3}) allows us to reduce the time
delay logarithm in equation (\ref{p6}) to another Lorentz-invariant form
\begin{equation}
\label{pp}
\ln\left(\frac{\rho_2-{\bm k}\cdot{\bm\rho}_2}{\rho_1-{\bm
k}\cdot{\bm\rho}_1}\right)=-\ln\left(\frac{\rho_{2R}+\rho_{1R}-\rho_{12}}{\rho_{2R}+\rho_{1R}+\rho_{12} }\right)\;,
\end{equation}
where the ranging distance $\rho_{12}=rk_\alpha u^\alpha= u_\alpha r^\alpha$
is invariant with respect to the Lorentz transformation. It represents
contraction of the null vector $r^\alpha$ defined in equation (\ref{p3})
with four-velocity $u^\alpha$ of the gravitating body. The null vector
$r^\alpha$ determines (unperturbed) propagation of the electromagnetic
signal. Distances $\rho_{1R}$, $\rho_{2R}$ are defined in equations
(\ref{p7}), (\ref{p8}), and they also represent contraction of the null
vectors $\rho^\alpha_1$, $\rho^\alpha_2$ with four-velocity $u^\alpha$ of
the gravitating body. However, contrary to vector $r^\alpha$, vectors
$\rho^\alpha_1$, $\rho^\alpha_2$ describe the null characteristics of the
gravitational field.

Accounting for equation (\ref{pp}) the Lorentz-invariant expression for the time delay assumes the following form
\begin{equation}
\label{qog}
\Delta t=\frac{2GM}{c^3}\frac{1-{\bm
k}\cdot{\bm\beta}}{\sqrt{1-\beta^2}}
\ln\left(\frac{\rho_{2R}+\rho_{1R}-\rho_{12}}{\rho_{2R}+\rho_{1R}+\rho_{12} }\right)\;.
\end{equation}
This equation is apparently Lorentz-invariant, valid for any value of the velocity of the light-ray deflecting body, and essentially generalizes the result of the paper by \citet{bai}.

\section{Post-Newtonian Expansion of the Ranging Delay}\label{pne}

Let us introduce an auxiliary vectors \citep{bd}
\begin{equation}
\label{x}
n^\alpha_2=\partial^\alpha \rho_{2R} = \frac{\rho^\alpha_2}{\rho_{2R}}-
u^\alpha\;,\qquad n^\alpha_1=\partial^\alpha \rho_{1R} =
\frac{\rho^\alpha_1}{\rho_{1R}}- u^\alpha\;
\end{equation}
Vectors $\rho^\alpha_2$ and $\rho^\alpha_1$ are null as defined by the
(gravity-field) null cone equations
(\ref{we1}), (\ref{we2}). The four-velocity of the body, $u^\alpha$, is a
time-like vector, $u_\alpha u^\alpha=-1$. The difference between the null
and time-like vector yields the space-like vectors $n^\alpha_2$,
$n^\alpha_1$, because
$n_{1\alpha} n^\alpha_1=n_{2\alpha} n^\alpha_2=+1$.

The post-Newtonian expansion of $z^\alpha(s_2)$ around
time $t_2$, and the post-Newtonian expansion of $z^\alpha(s_1)$ around
time $t_1$ are obtained by making use of a Taylor expansion. Omitting
acceleration, one gets
\begin{eqnarray}
\label{oo}
\rho^\alpha_2&=&r^\alpha_2-(s_2-t_2)\frac{dz^\alpha}{ds}=r^\alpha_2+\rho_2
u^\alpha\;,\\\label{oo1}
\rho^\alpha_1&=&r^\alpha_1-(s_1-t_1)\frac{dz^\alpha}{ds}=r^\alpha_1+\rho_1
u^\alpha\;,
\end{eqnarray}
and
\begin{eqnarray}
\label{vv}
\rho_2&=&\rho_{2R}+u_\beta r^\beta_2\;,\\\label{vv1}
\rho_1&=&\rho_{1R}+u_\beta r^\beta_1\;,
\end{eqnarray}
where the retarded time equations (\ref{we1}), (\ref{we2})
have been used to replace time intervals $s_2-t_2$ and $s_1-t_1$.
We have also introduced in previous equations the pure spatial vectors
\begin{eqnarray}
\label{e1}
r^\alpha_2&=&x^\alpha_2-z^\alpha(t_2)=\left\{r^0_2=0\;,\;r^i_2=x^i_2-z^i(t_2
)\right\}\;,\\
 \label{e2}
r^\alpha_1&=&x^\alpha_1-z^\alpha(t_1)=\left\{r^0_1=0\;,\;r^i_1=x^i_1-z^i(t_1
)\right\}\;,
\end{eqnarray}
which are lying on the hypersurface of constant time $t_2$ and $t_1$
respectively.

Substituting equations (\ref{oo})-- (\ref{vv1})
into equation (\ref{x}) reveals that
\begin{eqnarray}
\label{j1}
n^\alpha_2\rho_{2R}&=&r^\alpha_2+u^\alpha (u_\beta
r^\beta_2)\;,\\\label{j1i}
n^\alpha_1\rho_{1R}&=&r^\alpha_1+u^\alpha (u_\beta r^\beta_1)\;.
\end{eqnarray}
Taking into account that $n^\alpha_2$ and $n^\alpha_1$ are space-like unit
vectors, one has
\begin{eqnarray}
\label{poq}\rho_{2R}&=&\sqrt{r_{\alpha 1} r^\alpha_2+(u_\alpha
r^\alpha_2)^2}=\sqrt{\frac{r^2_2-({\bm\beta}\times{\bm
r}_2)^2}{1-\beta^2}}\;,\\\label{poq1}
\rho_{1R}&=&\sqrt{r_{\alpha 0} r^\alpha_1+(u_\alpha
r^\alpha_1)}=\sqrt{\frac{r^2_1-({\bm\beta}\times{\bm r}_1)^2}{1-\beta^2}}\;.
\end{eqnarray}
We further notice that, if acceleration is neglected,
\begin{equation}
\label{ppas}
\rho_{12}=\frac{{\bm k}\cdot{\bm\sigma}}{\sqrt{1-\beta^2}}\;r_{12}\;,
\end{equation}
where the unit vector
\begin{equation}
\label{p12}
{\bm\sigma}=\frac{{\bm k}-{\bm\beta}}{|{\bm k}-{\bm\beta}|}\;,
\end{equation}
the relative distance
\begin{equation}
\label{kai}
r_{12}=|{\bm r}_2-{\bm r}_1|\;,
\end{equation}
and
\begin{eqnarray}
\label{pp12}
{\bm r}_2&=&{\bm x}_2-{\bm z}(t_2)\;,\\
\label{pp13}
{\bm r}_1&=&{\bm x}_1-{\bm z}(t_1)\;,
 \end{eqnarray}
 are spatial distances from the observer to the body and from the emitter to
the body taken respectively at the time of reception and that of emission of
the electromagnetic wave. It is worth observing that the post-Newtonian expansion
of the Euclidean dot product ${\bm k}\cdot{\bm\sigma}$ does not have a term, which is linear with respect to velocity
\begin{equation}
\label{popk}
{\bm k}\cdot{\bm\sigma}=1-\frac12({\bm k}\times{\bm\beta})^2+O(\beta^3)\;.
\end{equation}
This expansion yields
\begin{equation}
\label{qmd}
\rho_{12}=r_{12}+O(\beta^2)\;,
\end{equation}
that is the distance $r_{12}$ is a Lorentz-invariant function up to the second
post-Newtonian corrections of the order of $\beta^2$. This justifies the replacement of the heliocentric coordinates of the massive bodies of the solar system to their barycentric counterparts introduced by Moyer in the ODP manual \citep{moy} ad hoc (see section \ref{nasa} of the present paper for further details).

After preceding preparations, we are ready to write down the post-Newtonian
expansion for the ranging time delay. We would like to emphasize that the
post-Newtonian expansion of the ranging delay is not unique and can be
represented in several different forms, which are physically and
computationally equivalent. However, this non-uniqueness complicates things
and has been debated in papers \citep{bai,kop09} regarding the nature of the relativistic time delay effects associated with motion of the gravitating body. In what follows, we derive
all possible forms of the post-Newtonian expansion of the ranging delay demonstrating that the relativistic effects associated with the motion of the light-ray deflecting body are induced by the gravitomagnetic field arising due to the translational motion of the body with respect to observer \citep{km,kop-cqg}.

First of all,
substituting equations (\ref{p7}), (\ref{p8}) to (\ref{pp}) casts the
ranging delay (\ref{tdz}) in the following form
\begin{equation}
\label{p11}
\Delta t=\frac{2GM}{c^3}\frac{1-{\bm
k}\cdot{\bm\beta}}{\sqrt{1-\beta^2}}\ln\left[\frac{\sqrt{r^2_2-({\bm\beta}\times{\bm r}_2)^2}+\sqrt{r^2_1-({\bm\beta}\times{\bm r}_1)^2}+({\bm
k}\cdot{\bm\sigma})r_{12}}{\sqrt{r^2_2-({\bm\beta}\times{\bm
r}_2)^2}+\sqrt{r^2_1-({\bm\beta}\times{\bm r}_1})^2-({\bm k}\cdot{\bm\sigma})r_{12}}\right]\;,
\end{equation}
which is the most convenient for making its explicit post-Newtonian
expansion with respect to the ratio of $\beta=v/c$. Neglecting terms of the
order of $\beta^3$ one has
\begin{eqnarray}
\label{p13}
\Delta t&=&\left(1-{\bm
k}\cdot{\bm\beta}+\frac12\beta^2\right)\frac{2GM}{c^3}\ln\left(\frac{r_1+r_2
+r_{12}}{r_1+r_2-r_{12}}\right)
\\\nonumber&+&
\frac{GM}{c^3}\frac{r_{12}}{r_1r_2}\frac{({\bm
n}_1\times{\bm\beta})^2r_1+({\bm n}_2\times{\bm\beta})^2r_2-({\bm
k}\times{\bm\beta})^2(r_1+r_2)}{1+{\bm n}_1\cdot{\bm
n}_2}+O\left(\frac{GM}{c^3}\beta^3\right)\;,
\end{eqnarray}
where the unit vectors ${\bm n}_1={\bm r}_1/r_1$, ${\bm n}_2={\bm r}_2/r_2$ with ${\bm r}_1$, ${\bm r}_2$ being defined in equations (\ref{pp12}), (\ref{pp13}) (see Fig. \ref{fig1}).

Velocity-dependent corrections appear in this expression {\it explicitly} as
the terms depending on ${\bm\beta}={\bm v}/c$, and {\it implicitly} in
the argument of the logarithm, which depends on two positions of the body
taken at times $t_1$ and $t_2$, that is ${\bm z}(t_2)={\bm z}(t_1)+{\bm
v}(t_2-t_1)={\bm z}(t_1)+{\bm \beta}r$ so that $r_2$ and $r_{12}$ are not independent of $r_1$. We discuss the impact of the
velocity-dependent terms on measured values of the PPN parameters in the
next section.

It is also instructive to derive the time delay equation in the linearized form as it is given in
\citet{will-livrev}. We
 make use of equations (\ref{x})-(\ref{poq1}) to get the post-Newtonian
expansion of functions entering the argument of the logarithm in the ranging
delay ({\ref{tde0})
\begin{eqnarray}
\label{aq1}
k_\alpha \rho^\alpha_2&=&k_\alpha r^\alpha_2+\left(k_\alpha
u^\alpha\right)\left[u_\beta r^\beta_2+\sqrt{r_{2\beta}
r^\beta_2+\left(u_\beta r^\beta_2\right)^2}\right]\;,\\
\label{aq1a}
k_\alpha \rho^\alpha_1&=&k_\alpha r^\alpha_1+\left(k_\alpha
u^\alpha\right)\left[u_\beta r^\beta_1+\sqrt{r_{1\beta}
r^\beta_1+\left(u_\beta r^\beta_1\right)^2}\right]\;.
\end{eqnarray}
Explicit expansion of these equations with respect to the powers of the
velocity-tracking parameter $\beta=v/c$ brings about the following result
\begin{eqnarray}
\label{pn1}
\rho_2-{\bm k}\cdot{\bm\rho}_2&=&r_2-{\bm k}\cdot{\bm
r}_2+{\bm\beta}\cdot{\bm r}_2-r_2\left({\bm
k}\cdot{\bm\beta}\right)+O\left(\beta^2\right)\;,\\
\label{pn2}
\rho_1-{\bm k}\cdot{\bm\rho}_1&=&r_1-{\bm k}\cdot{\bm
r}_1+{\bm\beta}\cdot{\bm r}_1-r_1\left({\bm
k}\cdot{\bm\beta}\right)+O\left(\beta^2\right)\;.
\end{eqnarray}
Applying these expansions to the argument of logarithm in the ranging delay
(\ref{tdz}) yields the first term in the post-Newtonian expansion of the ranging delay in the
form given in \citep{will-livrev}
\begin{equation}
\label{p16}
\Delta t=(1-{\bm
k}\cdot{\bm\beta})\frac{2GM}{c^3}\ln\left[\frac{r_2-{\bm\sigma} \cdot{\bm
r}_2}{r_1-{\bm\sigma} \cdot{\bm
r}_1}\right]+O\left(\frac{2GM\beta^2}{c^3}\right)\;,
\end{equation}
where the unit vector
\begin{equation}
\label{p17}
{\bm\sigma}={\bm k}-{\bm k}\times({\bm\beta}\times{\bm k})+O(\beta^2)\;,
\end{equation}
 is the same as that defined by equation (\ref{p12}).

The explicit
post-Newtonian dependence of the time delay on velocity of the gravitating
body ${\bm v}$ enters the argument of the logarithm in the form of equation
(\ref{p17}), which looks like the aberration of light for the unit vector
${\bm k}$.  However, equation (\ref{p16}) approximates
the exact time delay equation (\ref{tde}), which demonstrates that the
argument of the logarithmic function is a 4-dimensional dot product
$k_\alpha\rho^\alpha$ of two null vectors $k^\alpha$ and $\rho^\alpha$.
Vector $k^\alpha$ points out the direction of propagation of light ray, while
the null vector $\rho^\alpha=x^\alpha-z^\alpha(s)$ points out the direction of
the null characteristic of the gravity field equations. The Lorentz
transformation, $\Lambda^{\alpha'}{}_\beta$, from one frame to another changes the null vector $k^{\alpha
'}=\Lambda^{\alpha '}{}_\beta k^\beta$, but in order to preserve the
Lorentz-invariance of the gravitational time delay $\Delta t$, the null vector
$\rho^\alpha$ directed along the body's gravity field must change accordingly
$\rho^{\alpha '}=\Lambda^{\alpha '}{}_\beta \rho^\beta$, so that the dot product
$k_\alpha\rho^\alpha=k_{\alpha '}\rho^{\alpha '}$ remains the same. Hence, not only the light undergoes aberration, when one goes from one frame to another, but the null characteristics of the gravitational field in the time delay $\Delta t$ must change too in the same proportion, if general relativity is valid. In other words, equation (\ref{p17}) is not the ordinary equation of the aberration of light in flat space-time (without gravity field) but a more profound relationship for a curved space-time showing that even in the presence of the gravitational field of the moving body, affecting the light propagation, the aberration of light equation remains the same as in the flat space-time. This can be true if and only if both the gravitational field perturbation $h_{\alpha\beta}$ and the affine connection $\Gamma^\alpha_{\beta\gamma}$ remain invariant under the Lorentz group transformation, which is parameterized with the same fundamental speed $c$ as the Lorentz group of the underlying electromagnetic wave used in the ranging time-delay experiment. This interpretation is further discussed in more detail elsewhere \citep{kffp,kop-fom,kmak}.

\section{Coupling of the PPN Parameters with the Velocity-Dependent Terms}
\subsection{Explicit Coupling}
Equation (\ref{p13}) describes the Lorentz transformation of the (static)
Shapiro time delay from the rest frame of the massive body (Sun, planet) to
the frame of reference in which the data processing is performed. For we have restricted ourselves with the post-Newtonian expansion of the linearized time delay up to the terms which are quadratic with respect to velocity of the moving gravitating body, equation (\ref{p13}) can be superimposed with the static terms
of the second order with respect to the universal gravitational constant $G$
entering equation (\ref{aq4}). This is because these terms have the same order of magnitude so that we do not need to develop the Lorentz invariant expression for the terms which are quadratic with respect to $G$. We shall also neglect for simplicity the terms which are products of $\beta^2$ with the PPN parameter $\bar\gamma_{PPN}$ because $\bar\gamma_{PPN}$ has been already limited by the solar system experiments up to the value not exceeding $10^{-4}$. Thus, the product $\bar\gamma_{PPN}\beta^2$ exceeds the accuracy of the post-post-Newtonian approximation.

Our calculation yields the following, Lorentz-invariant equation for the
post-post-Newtonian time delay
\begin{eqnarray}
\label{p13a}
\Delta t&=&\left(1+\frac{\bar\gamma_{PPN}}{2}-{\bm
k}\cdot{\bm\beta}-\frac{\bar\gamma_{PPN}}{2}{\bm
k}\cdot{\bm\beta}+\frac12\beta^2\right)\frac{2GM}{c^3}\ln\left(\frac{r_1+r_2
+r_{12}}{r_1+r_2-r_{12}}\right)
\\\nonumber&+&\left(1+\frac{\bar\gamma_{PPN}}{2}\right)\frac{GM}{c^3}\frac{r_{12}}{r
_1r_2}\frac{({\bm n}_1\times{\bm\beta})^2r_1+({\bm
n}_2\times{\bm\beta})^2r_2-({\bm k}\times{\bm\beta})^2(r_1+r_2)}{1+{\bm
n}_1\cdot{\bm n}_2}\\\nonumber&+&
\frac{G^2M^2}{c^5}\frac{r_{12}}{r_1r_2}\left[\left(\frac{15}{4}+2\bar\gamma_{PPN}-
\bar\beta_{PPN}+\frac{3}{4}\bar\delta_{PPN}\right)\frac{\arccos({\bm n}_1\cdot{\bm
n}_2)}{|{\bm n}_1\times{\bm n}_2|}-\frac{(2+\bar\gamma_{PPN})^2}{1+{\bm
n}_1\cdot{\bm n}_2}\right]
+O\left(\frac{GM}{c^3}\beta^3\right)\;.
\end{eqnarray}
One can immediately observe that the PPN parameter $\bar\gamma_{PPN}$ couples with the
velocity terms in front of the logarithmic term. This means that the
amplitude of the Shapiro delay is effectively sensitive to the linear
combination
\begin{equation}
\label{por5}
\bar\Gamma=\bar\gamma_{PPN}-2\beta_R-2\bar\gamma_{PPN}\beta_R+\beta^2_R+\beta^2_T\;,
\end{equation}
that will be measured in high-precision space-based experiments like
BepiColombo, ASTROD, LATOR, etc. Here and elsewhere, we denote
respectively $\beta_R\equiv{\bm k}\cdot{\bm\beta}$ -- the radial velocity, and $\beta_T\equiv|{\bm
k}\times{\bm\beta}|$  -- the transverse
velocity of the massive body that deflects the light ray.

Equation (\ref{por5}) elucidates that the measured value $\bar\Gamma$ of the
parameter $\bar\gamma_{PPN}$ is affected by the velocity terms, which explicitly
present in the post-Newtonian expansion of the Shapiro time delay. In case
of the ranging gravitational experiment in the field of Sun with the light
ray grazing the solar limb, one has $d=R_\odot=7\times 10^{10}$ cm -- the
solar radius, and $r_g=3\times 10^5$ cm -- the Schwarzschild radius of the
Sun. The Sun, in moving in its orbit around the barycenter, has an average
distance of 1.1 $R_\odot$ from it but may be as far as 2.3 $R_\odot$. The orbital
path of the Sun about the barycenter traces out a curve that is closely
resemble an epitrochoid -- three-lobed rosette, with three large and three
small loops -- with a loop period of 9 to 14 years. Fifteen successive
orbits comprise a 179-year cycle of the solar motion around the barycenter
\citep{1987SoPh..110..191F,1965AJ.....70..193J} -- the duration, which is
also the time taken for the planets to
occupy approximately the same positions again relative to each
other and the Sun. The solar velocity $v_\odot$ with respect to the
barycenter of the solar system can reach maximal value of 15.8 m/s giving
rise to $\beta_\odot=v_\odot/c=5.3\times 10^{-8}$. Because space missions LATOR and ASTROD are going to measure $\bar\gamma_{PPN}$ parameter with a precision
approaching to $10^{-9}$ \citep{2004CQGra..21.2773T,2007NuPhS.166..153N}, the explicit velocity-dependent correction to the
Shapiro time delay in the solar gravitational field must be apparently taken into account. Current
indeterminacy in the solar velocity vector is about 0.366 m/day
\citep{pitjeva} that yields an error of $\Delta\beta_\odot\simeq 1.4\times
10^{-14}$. This error is comparable with the contribution of the
second-order velocity terms $\beta^2_\odot\le 2.8\times 10^{-15}$. However, they are too small and can
be neglected in the measurement of $\bar\gamma_{PPN}$.

Coupling of the velocity-dependent terms with parameters $\bar\beta_{PPN}$ and
$\bar\delta_{PPN}$ can be understood after making expansion of high-order terms in
equation (\ref{p13a}) with respect to the impact parameter of the light ray
$d=|{\bm k}\times{\bm r}_1|=|{\bm k}\times{\bm r}_1|$ that is assumed to be
small: $d\ll r_1$, $d\ll r_2$. The unit vectors ${\bm n}_1$ and ${\bm n}_2$ can be decomposed in the
post-post-Newtonian terms as follows
\begin{eqnarray}
\label{pb7}
{\bm n}_1&=&-{\bm k}\cos\theta_1+{\bm n}\sin\theta_1\;,\\\label{pb8}
{\bm n}_2&=&\;\;\,{\bm k}\cos\theta_2+{\bm n}\sin\theta_2\;,
\end{eqnarray}
where the unit vector ${\bm n}$ is directed from the massive body to the
light-ray trajectory along the impact parameter: ${\bm d}=d{\bm n}$. It is
convenient to introduce the deflection angle $\theta$ defined as
\begin{equation}
\label{aj5}
{\bm n}_1\cdot{\bm n}_2=\cos(\pi-\theta)=-\cos\theta\;.
\end{equation}
One can easily observe that $\theta=\theta_1+\theta_2$. Practically all
gravitational ranging experiments are done in the small-angle approximation,
when $\theta\ll 1$, $\theta_1\ll 1$, $\theta_2\ll 1$. In this approximation,
one has
\begin{eqnarray}
\label{aj6}
1+{\bm n}_1\cdot{\bm
n}_2&=&\frac{\theta^2}{2}+O\left(\theta^4\right)\;,\\\label{aj7}
({\bm n}_1\times{\bm\beta})^2r_1+({\bm n}_2\times{\bm\beta})^2r_2-({\bm
k}\times{\bm\beta})^2(r_1+r_2)&=&\theta
d\left(\beta^2_R-\beta^2_T\right)+O\left(\theta^3\right)\;,
\end{eqnarray}
Substituting equations (\ref{por5}), (\ref{aj5})--(\ref{aj7}) to equation
(\ref{p13a}) yields
\begin{eqnarray}
\label{w13a}
\Delta
t&=&\left(2+\bar\Gamma\right)\frac{GM}{c^3}\ln\left(\frac{r_1+r_2+r_{12}}{r_1+
r_2-r_{12}}\right)
\\\nonumber
&+&
\frac{G^2M^2}{c^5}\frac{r_{12}}{r_1r_2}\left[\left(\frac{15}{4}+2\bar\gamma_{PPN}-
\bar\beta_{PPN}+
\frac{3}{4}\tilde\delta_{PPN}\right)\frac{\pi}{\theta}-\frac{2(2+\bar\gamma_{PPN})^2}{\theta^2}\right]
+O\left(\frac{GM}{c^3}\beta^3\right)\;,
\end{eqnarray}
where we have introduced a new notation
\begin{equation}
\label{w14a}
\tilde\delta_{PPN}\equiv\bar\delta_{PPN}+\left(1+\frac{\bar\gamma_{PPN}}{2}\right)
\frac{16}{3\pi}\frac{d}{r_g}\left(\beta^2_R-\beta^2_T\right)\;,
\end{equation}
and denoted $r_g\equiv 2GM/c^2$ -- the Schwarzschild radius of the massive body
deflecting the light ray.
Explicit contribution of the solar velocity terms to the parameter
$\tilde\delta_{PPN}$ can achieve $1.1\times 10^{-9}$ that is much less than the
precision of measurement of the PPN parameter $\bar\delta_{PPN}$ in LATOR and
ASTROD missions \citep{2006CQGra..23..309P} and can be currently neglected.

We recall to the reader that in scalar-tensor theory of gravity parameter $\bar\beta_{PPN}$ can not be determined separately from $\bar\delta_{PPN}$ as they appear in the linear combination $-\bar\beta_{PPN}+3/4\bar\delta_{PPN}$. Following \citep{2006CQGra..23..309P} we assume that $\bar\beta_{PPN}$ is determined from other kind of gravitational experiments, and eliminate it from the fitting procedure.

\subsection{Implicit Coupling}\label{icoup}

In the previous section we have made an explicit post-Newtonian expansion of the
ranging time delay in powers of the velocity-tracking parameter $\beta=v/c$. This
post-Newtonian expansion is shown in equation (\ref{p13a}). It looks like
the only place, where the linear velocity correction to the Shapiro delay appears, is in front of the logarithmic term. However, a scrutiny
analysis reveals that the linear velocity-dependent correction is also present {\it implicitly} in the argument of
the logarithmic function. Indeed, distances $r_1=|{\bm x}_1-{\bm z}(t_1)|$
and $r_2=|{\bm x}_1-{\bm z}(t_2)|$ depend on two positions of the massive
body taken at two different instants of time, $t_1$ and $t_2$. The body
moves as light propagates from the point of emission ${\bm x}_1$ to the
point of observation ${\bm x}_2$, so that the coordinates of the body are not arbitrary but connected through a relationship
\begin{equation}
\label{bxc3}
{\bm z}(t_2)={\bm z}(t_1)+{\bm v}(t_2-t_1)\;,
\end{equation}
which, indeed, shows that the velocity of the body is involved in calculation of the numerical value of the argument of the time-delay logarithm.

Though this dependence on the velocity of the massive body is implicit, it definitely affects the measured values of the PPN parameters  and makes their values biased in case if either general relativity is invalid or if the numerical code used for data processing of the ranging experiment, does not incorporate the solar system ephemeris properly \citep{kpsv}. Let us show how this impact on the PPN parameters can
happen.

To this end we shall assume that the light ray passes at a minimal
distance $d$ from the body at the time of the closest approach $t_*$ which
is defined in the approximation of the unperturbed light-ray trajectory,
${\bm x}(t)={\bm x}_1+{\bm k}(t-t_1)$ for $t\ge t_1$ or ${\bm x}(t)={\bm
x}_2+{\bm k}(t-t_2)$ for $t\le t_2$, from the condition \citep{klikop}
\begin{equation}
\label{bcx5}
\left\{\frac{d|{\bm x}(t)-{\bm z}(t)|}{dt}\right\}_{t=t_*}=0\;,
\end{equation}
where ${\bm x}(t)={\bm x}_1+{\bm k}(t-t_1)$ is the (unperturbed) light-ray trajectory, and ${\bm z}(t)={\bm z}(t_1)+{\bm v}(t-t_1)$ is the body's world line in the approximation of a straight line, uniform motion.
Taking the time derivative and solving the equation yield
\begin{equation}
\label{bcx6}
t_*=t_1-\frac{{\bm\sigma}\cdot{\bm r}_1}{c|{\bm
k}-{\bm\beta}|}=t_2-\frac{{\bm\sigma}\cdot{\bm r}_2}{c|{\bm
k}-{\bm\beta}|}\;,
\end{equation}
where the unit vector ${\bm\sigma}$ has been defined in equation
(\ref{p12}). The post-Newtonian expansion of various distances near the time of the
closest approach gives us
\begin{eqnarray}
\label{bc1}
r_1&=&r_{1*}\left[1-\left({\bm\beta}\cdot{\bm
n}_{1*}\right)\frac{l_1}{r_{1*}}+\frac{\left({\bm\beta}\times{\bm
n}_{1*}\right)^2}{2}\left(\frac{l_1}{r_{1*}}\right)^2\right]\;,\\\label{bc2}
r_2&=&r_{2*}\left[1-\left({\bm\beta}\cdot{\bm
n}_{2*}\right)\frac{l_2}{r_{2*}}+\frac{\left({\bm\beta}\times{\bm
n}_{2*}\right)^2}{2}\left(\frac{l_2}{r_{2*}}\right)^2\right]\;,\\\label{bc3}
r_{12}&=&\;\;r_{\;\:}\left[1-{\bm\beta}\cdot{\bm
k}+\frac{\left({\bm\beta}\times{\bm k}\right)^2}{2}\right]\;,
\end{eqnarray}
where $l_1=c(t_1-t_*)$, $l_2=c(t_2-t_*)$, the unit vectors ${\bm n}_{1*}={\bm r}_{1*}/r_{1*}$, ${\bm n}_{2*}={\bm r}_{2*}/r_{2*}$, and distances ${\bm r}_{1*}={\bm x}_1-{\bm z}(t_{*})$, ${\bm r}_{2*}={\bm x}_2-{\bm z}(t_{*})$.

We substitute now the post-Newtonian expansions (\ref{bc1})--(\ref{bc3}) to
the logarithmic function of the Shapiro time delay and apply the small-angle
approximation. It will yield
\begin{equation}
\label{bc4}
\ln\left(\frac{r_1+r_2+r_{12}}{r_1+r_2-r_{12}}\right)=
\ln\left(\frac{r_{1*}+r_{2*}+r}{r_{1*}+r_{2*}-r}\right)-
\frac{2rd_*}{r_{1*}r_{2*}}\frac{{\bm
k}\cdot{\bm\beta}}{\theta_*}\Bigl[1+O\left(\beta\right)+O\left(\theta_*\right)\Bigr]\;,
\end{equation}
where $\theta_*$ is the angle between two vectors ${\bm n}_{1*}$ and ${\bm
n}_{2*}$ defined as ${\bm n}_{1*}\cdot{\bm n}_{2*}=\cos(\pi-\theta_*)$.

The post-Newtonian expansion of the ranging delay in the vicinity of the
time of the closest approach of the light ray to the massive body reveals
that the parameter $\bar\delta_{PPN}$ is affected by the first-order velocity terms
from equation (\ref{bc4}). Specifically, taking into account equation
(\ref{bc4}) allows us to write down the ranging delay in the following form
\begin{eqnarray}
\label{w13c}
\Delta
t&=&\left(2+\bar\Gamma\right)\frac{GM}{c^3}\ln\left(\frac{r_{1*}+r_{2*}+r}{r_{
1*}+r_{2*}-r}\right)
\\\nonumber&+&
\frac{G^2M^2}{c^5}\frac{r}{r_{1*}r_{2*}}\left[\left(\frac{15}{4}+2\bar\gamma_{PPN}
-\bar\beta_{PPN}+
\frac{3}{4}\bar\Delta\right)\frac{\pi}{\theta_*}-\frac{2(2+\bar\gamma_{PPN})^2}{\theta^2_*}\right]
+O\left(\frac{GM}{c^3}\beta^3\right)\;,
\end{eqnarray}
where
\begin{equation}
\label{w14b}
\bar\Delta\equiv\bar\delta_{PPN}-\left(1+\frac{\bar\gamma_{PPN}}{2}\right)
\frac{16}{3\pi}\frac{d}{r_g}\;\beta_R\;.
\end{equation}
The last term in equation (\ref{w14b}) can amount to 0.02, which exceeds the
expected accuracy of measuring the PPN parameter $\bar\delta_{PPN}$ with
LATOR/ASTROD missions by a factor of 10 as follows from \citep{2006CQGra..23..309P}. This
clearly indicates the necessity of inclusion of the velocity-dependent
post-Newtonian corrections to the data analysis of the high-precise time
delay and ranging gravitational experiments.

\section{Ranging Experiments and Lorentz Invariance of Gravity.}

In special relativity, where the Minkowski geometry represents a flat space-time, the Lorentz symmetry is a global symmetry consisting of
rotations and boosts. However, in curved space-time, in the most general
case, the Lorentz symmetry is a local symmetry that transforms local vectors and
tensors in the tangent (co-tangent) space at each space-time point. Nonetheless, general
relativity admits the global Lorentz symmetry, at least, for isolated astronomical
systems residing in asymptotically-flat space-time \citep{fock}. This asymptotic
Lorentz symmetry of gravitational field can be traced in the invariant
nature of the gravitational Li\'enard-Wiechert potentials given by equation (\ref{1}), which are
solutions of the linearized Einstein equations. The asymptotic
Minkowskian space-time for isolated systems defines the background manifold for gravitational field perturbations, $h_{\alpha\beta}$, and must have the same null-cone structure
as the local tangent space-time, which is defined by motion of light particles (photons). However, this theoretical argument is a matter of experimental study \citep{kop2001,fk-apj}.

Ranging time-delay experiments are, perhaps, the best experimental technique for making
such test. This is because they operate with the gauge-invariant fundamental
field of the Maxwell theory having well-established and unambiguous physical
properties. Propagation of radio (light) signals traces the local structure of the null cone hypersurface all the way
from the point of emission down to the point of its observation. Now, if the massive
body, which deflects radio (light) signals, is static with respect to observer, one can not draw any conclusion on
the asymptotic structure of the space-time manifold and on whether its Lorentz symmetry is compatible with the Lorentz symmetry of the light cone. This is because the gravitational
interaction of the body with the radio (light) signal is realized in the form of the
instantaneous Coulomb-like gravitational force with having no time derivatives of the gravitational potentials been involved. However, if the massive body
is moving with respect to observer as light propagates, its gravitational force is not instantaneous
and must propagate on the hypersurface of the null cone of the asymptotic Minkowskian space-time as it is
described by the Li\'enard-Wiechert gravitational potentials (\ref{1}). The terms in the ranging time-delay (\ref{tdz})
depending on both the translational velocity ${\bm \beta}={\bm v}/c$ of the massive body and the retarded time $s$, originate from the time derivatives of the gravitational potentials and characterize the global Lorentz symmetry of the gravitational field. Therefore, measurement of these terms in the ranging time-delay experiments has a fundamental significance \citep{kffp}.

Currently, there is a growing interest of theoretical physicists to
gravitational theories where the global Lorentz symmetry of gravitational field can be
spontaneously violated \citep{rb}. This is motivated by the need of unification
of the gravity field with other fundamental interactions. These theories
introduce additional long-range fields to the gravitational Lagrangian,
which destroy the symmetry between the, so-called, observer and particle
invariance \citep{kos1,kos2,kos3}.
Interaction terms involving these fields appear also
in the equations of motion of test particles.
It is the interaction with these fields that can lead to physical effects of
the broken Lorentz symmetry that can be tested in experiments. Outcome of
these experiments depends crucially on the assumptions made about the
structure of the additional terms in the gravitational Lagrangian and the
numerical value of the coupling constants of these fields with matter. On the other hand, the measurement of the post-Newtonian velocity-dependent and/or retarded-time
corrections in the ranging time-delay experiments does not depend on any additional
assumptions and relies solely on general relativistic prediction of how the radio (light) signals
propagate in time-dependent gravitational fields.

It is remarkable that current technology already allows us to measure the
velocity-dependent and/or retarded-time post-Newtonian corrections in the ranging time-delay
experiments conducted in the solar system. The most notable experiment had
been done in 2002 with the VLBI technique \citep{fk-apj}. It measured the
retarded component of the near-zone gravitational field of Jupiter via its impact on the magnitude of the deflection angle of light from a
quasar \citep{kop2001,kop-cqg}. \citet{fom09} have repeated this {\it retardation of gravity} experiment in 2009 by making use of the close encounter of Jupiter and Saturn with quasars in the plane of the sky.

The Cassini experiment \citep{b0,a04} is also sensitive to the time-dependent perturbation of gravitational field of the Sun caused by its orbital motion around the barycenter of the solar system \citep{kpsv,bai,kop09}. However, its detection requires re-processing of the Cassini data in order to separate the Cassini measurement of PPN parameter $\gamma_{PPN}$ from the gravitomagnetic deflection of light by the moving Sun \citep{kpsv,kop09}.

\section{Ranging Delay in the NASA Orbit Determination Program}\label{nasa}

Relativistic ranging time delay, incorporated to the  NASA ODP code, was
originally calculated by \citet{moy} under assumption that the
gravitating body that deflects light, does not move. Regarding the Sun, it
means that the ODP code derives the ranging delay in the heliocentric frame.
Let us introduce the heliocentric coordinates $X^\alpha=(cT,X^i)$, and use
notation $x^\alpha=(ct, x^i)$ for the barycentric coordinates of the solar
system, which origin is at the center of mass of the solar system. The Sun
moves with respect to the barycentric frame with velocity ${\bm
v}_\odot=d{\bm x}_\odot/dt$ amounting to $\sim 15$ m/s. Though this velocity
looks small, it can not be neglected in such high-precision relativity
experiments as, for example, Cassini \citep{kpsv}.
A legitimate question arises whether the ODP code accounts for the solar
motion or not. We demonstrate in this appendix that the ranging time delay
in the ODP code is consistent with general relativity in the linear-velocity
approximation, but it fails to take into account the quadratic velocity
terms properly. Thus, more advanced theoretical development of the ODP code is required.

The ranging time delay in the heliocentric coordinates with the Sun located
at the origin of this frame, follows directly from equation (\ref{qog}) after
making use of the heliocentric coordinates. It reads
\begin{eqnarray}
\label{tdhf}
T_2-T_1&=&\frac{1}{c}|{\bm X}_2-{\bm X}_1|+\Delta T\;,\\\label{ok}
\Delta
T&=&\frac{2GM_\odot}{c^3}\ln\left[\frac{R_2+R_1+R_{12}}{R_2+R_1-R_{12}}\right]\;,
\end{eqnarray}
where ${\bm X}_2$ and ${\bm X}_1$ are the heliocentric coordinates of
observer and emitter respectively, distance of the emitter from the Sun is
$R_2=|{\bm X}_2|$, distance of the observer from the Sun is $R_1={\bm X}_1$,
and $R_{12}=|{\bm X}_2-{\bm X}_1|$ is the null heliocentric distance between
the emitter and observer. This equation coincides exactly (after reconciling
our and Moyer's notations for distances) with the ODP time-delay equation
(8-38) given in section 8 of the ODP manual \citep{moy} on page 8-19.
\citet{moy} had transformed the argument of the logarithm in the heliocentric
ranging delay (\ref{ok}) to the barycentric frame by making use of
substitutions
\begin{equation}
\label{jk}
{\bm X}_2\Rightarrow {\bm r}_2={\bm x}_2-{\bm x}_\odot(t_2)\qquad\;,\qquad {\bm X}_1\Rightarrow {\bm r}_2={\bm
x}_1-{\bm x}_\odot(t_1)\;.
\end{equation}
The ODP manual \citep{moy} does not provide any evidence that these
substitutions in the ranging time delay (\ref{ok}) are consistent with general relativity and do not violate the Lorentz symmetry. Nonetheless,
comparison of equations (\ref{ok}), (\ref{jk}) with the post-Newtonian
expression (\ref{p13}) for the ranging delay demonstrates that equations
(\ref{jk}) are legitimate transformations from the heliocentric to the
barycentric frame in the sense that they take into account velocity of the
Sun in the ranging time delay in the linearized, post-Newtonian term following the static Shapiro time delay.

Equation (\ref{p13}) also shows that the ODP code is missing the
velocity-dependent term in front of the logarithmic function in equation
(\ref{ok}). The ranging time delay in the heliocentric and barycentric
frames must be related by the simple equation
\begin{equation}
\label{bm}
\Delta t=(1-{\bm k}\cdot{\bm\beta}_\odot) \Delta T\;,
\end{equation}
which is a linearized version of equation (\ref{p13}) that was derived by \citet{ks}. We conclude that the ODP code used by NASA for navigation of spacecrafts in deep space, is missing a high-order velocity-dependent corrections to the Shapiro time delay and can not be used for processing and unambiguous interpretation of near-future ranging experiments in the solar system. A corresponding relativistic modification and re-parametrization of the ODP code based on equations of the present paper is highly required.

Equation (\ref{bm}) has been also derived by \citet{bai} who
claimed that the velocity-dependent terms appear in the time delay only in
front of the logarithmic function in equation (\ref{bm}). As we have shown in section \ref{icoup} the argument of the
logarithm in equation (\ref{p11}) also contains terms depending on velocity
${\bm v}$ of the gravitating body, which are {\it implicitly} present in the
definition of the distance $r_{12}$. This distance is calculated between two
 spatial points separated by the time interval required by light to travel between the point of emission and observation respectively (see equations (56)--(58)). Coordinates ${\bm z}(t_1)$ and ${\bm z}(t_2)$ are not the same because the gravitating body is moving. These coordinates are related by means of the equation (72), which demonstrates that velocity ${\bm v}$ of the gravitating body must be known in order to calculate the distance $r_{12}$. Because one has to rely upon equation (72) in the ODP data processing algorithm, the post-Newtonian expansion of distance $r_{12}$ yields
\begin{equation}
\label{p14q}
r_{12}=r-{\bm r}\cdot{\bm\beta}+O(\beta^2)\;,
\end{equation}
where the null distance $r=|{\bm r}|$ is defined in equation (\ref{p3}). It
follows that the distances $r_{12}$ and $r$ entering equation
(\ref{p14q}) are not the same quantities as they differ by terms of the
order of $v/c$. Equation (\ref{p14q}) reduces the ranging delay (\ref{bm})
to the following form
\begin{equation}
\label{p15}
\Delta t=(1-{\bm
k}\cdot{\bm\beta})\frac{2GM}{c^3}\ln\left[\frac{r_2+r_1+r-{\bm
r}\cdot{\bm\beta}}{r_2+r_1-r+{\bm
r}\cdot{\bm\beta}}\right]+O\left(\frac{2GM\beta^2}{c^3}\right)\;,
\end{equation}
which has been derived in our paper \citep{kpsv}.
\citet{bai} claimed that the expression (\ref{p15}) for the
ranging time delay does not appear in the ODP manual \citep{moy} and is not allowed for
theoretical analysis of the Cassini experiment as we did in \citep{kpsv}. However, expression (\ref{p15}) is exactly the same
function $\Delta T$ given in the ODP manual but expressed, instead of distance $r_{12}$, in terms of the
distance $r$ and velocity of the Sun, ${\bm v}$, via self-consistent mathematical transformation (\ref{p14q}).
For this reason, the two expressions are mathematically equivalent and either of them can be used in data processing of the ranging observations of the Cassini experiment \citep{kop09}.

\section*{Acknowledgments}
This work was supported by the Research Council Grant No. C1669103 of the University of Missouri-Columbia. I am grateful to Dr. Slava Turyshev (JPL) for valuable conversations and critical comments, which helped to improve this paper.

\newpage
\begin{figure}
\begin{center}
\includegraphics[width=20cm]{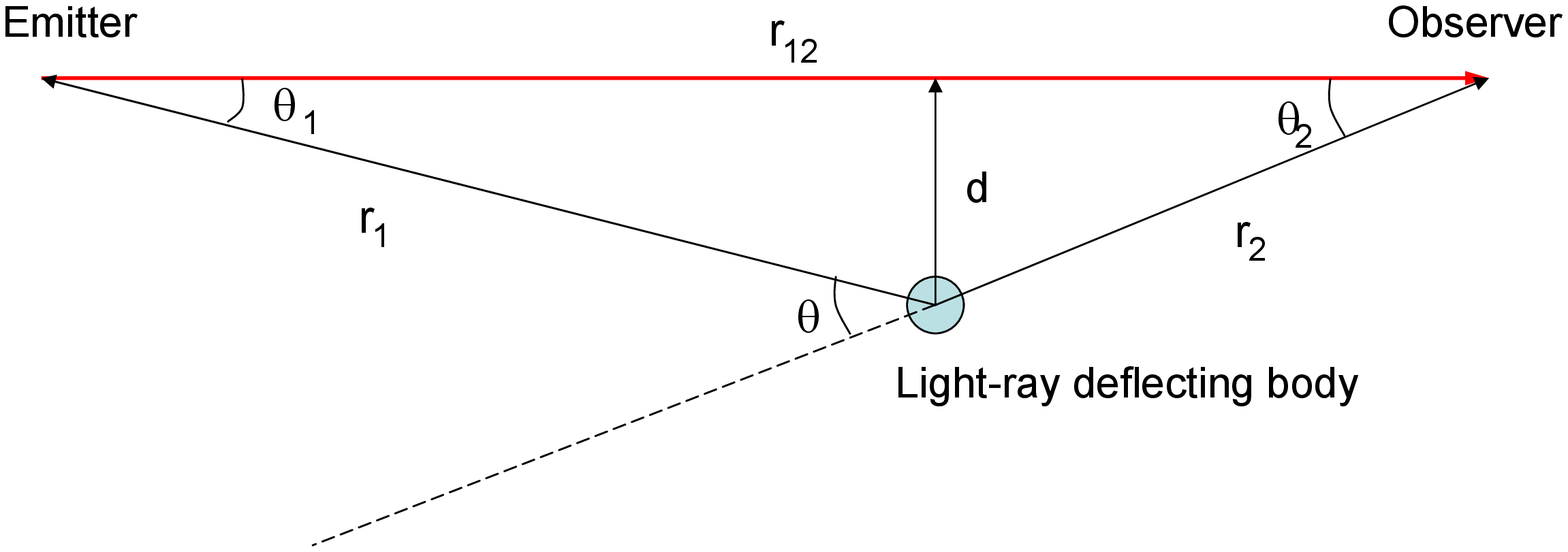}
\caption{Ranging time delay experiment. Electromagnetic signal is emitted at distance $r_1$ from the massive body, passes by it at the minimal distance d, and is received by observer at distance $r_2$. The emitter, observer, and the massive body move with respect to each other as the electromagnetic signal propagates. This makes the ranging delay experiment sensitive to the null cone structure of space-time in general relativity and modifies the Shapiro time delay.}
\label{fig1}
\end{center}
\end{figure}
\bsp
\label{lastpage}

\begin{thebibliography}{99}
\bibitem[\protect\citeauthoryear{Anderson, Lau \& Giampieri}{2004}]{a04}Anderson, J.~D., Lau, E.~L.
\& Giampieri, G., 2004, in: Proc. of the 22nd Texas Symp. on Rel. Astrophys.,
Stanford, eConf C041213, 0305 (http://www.slac.stanford.edu/econf/C041213/papers/0305.PDF)
\bibitem[\protect\citeauthoryear{Bel et al.}{1981}]{bd} Bel, L.,  Deruelle, N.,
Damour, T., Ibanez, J. \& Martin, J., 1981, {\it Gen. Rel. Grav.} {\bf 13}, 963
\bibitem[\protect\citeauthoryear{Bertotti, Iess \& Tortora}{ 2003}]{b0} Bertotti, B., Iess, L. \& Tortora, P., 2003, {\it Nature}, {\bf 425}, 374
\bibitem[\protect\citeauthoryear{Bertotti, Ashby \& Iess}{2008}]{bai} Bertotti, B., Ashby, N. \& Iess, L., 2008, {\it Class. Quantum Grav.}, {\bf 25}, 045013
\bibitem[\protect\citeauthoryear{Bluhm}{2008}]{rb} Bluhm, R., \ 2008, {\it Effects of Spontaneous Lorentz Violation in
Gravity}, arXiv:0801.0141
\bibitem[\protect\citeauthoryear{Blanchet, Faye \& Ponsot}{1998}]{blanch}Blanchet, L., Faye, G. \& Ponsot, B., \ 1998, \prd, {\bf 58}, 124002
\bibitem[\protect\citeauthoryear{Brumberg} {1992}]{brum} Brumberg, V.A., 1992, {\it Essential Relativistic
Celestial Mechanics}, (Adam Hilger: Bristol)
\bibitem[\protect\citeauthoryear{Colladay \& Kosteleck\'y}{1997}]{kos2} Colladay, D. \& Kosteleck\'y, V.~A., \ 1997, \prd, {\bf  55}, 6760
\bibitem[\protect\citeauthoryear{Colladay \& Kosteleck\'y}{1998}]{kos3} Colladay, D. \& Kosteleck\'y, V.~A., \ 1998, \prd, {\bf  58}, 116002
\bibitem[\protect\citeauthoryear{Damour \& Esposito-Far\`ese}{1992}]{sttg}Damour, T. \&
Esposito-Far\`ese, G., 1992, {\it Class. Quantum Grav.}, {\bf 9}, 2093
\bibitem[\protect\citeauthoryear{Damour \& Esposito-Far\`ese}{ 1996}]{dametal} Damour, T. \& Esposito-Far\`ese, G., 1996, \prd, {\bf 3}, 5541
\bibitem[\protect\citeauthoryear{Deng, Xie
\& Huang}{2009}]{xie} Deng, X.-M., Xie, Y.,
\& Huang, T.-Y.,\ 2009, \prd, {\bf 79}, 044014
\bibitem[\protect\citeauthoryear{Fairbridge \& Shirley}{1987}]{1987SoPh..110..191F} Fairbridge, R.~W. \& Shirley, J.~H.,\
1987, \solphys, {\bf 110}, 191
\bibitem[\protect\citeauthoryear{Fock}{1964}]{fock} Fock, V.~A.,\ 1964, {\it The Theory of Space, Time and Gravitation}, (The Macmillan Company: New York)
\bibitem[\protect\citeauthoryear{Fomalont \& Kopeikin}{2003}]{fk-apj} Fomalont, E.~B. \& Kopeikin, S.~M., 2003, {\it Astrophys. J.}, {\bf 598}, 704
\bibitem[\protect\citeauthoryear{Fomalont et al.}{2009}]{fom09} Fomalont, E.~B.,
Kopeikin, S., Titov, O.,
\& Honma, M.,\ 2009, American Astronomical Society, IAU Symposium \#261.~ Relativity in Fundamental Astronomy: Dynamics, Reference Frames, and Data Analysis 27 April - 1 May 2009 Virginia Beach, VA, USA, \#15.03, 261, 1503
\bibitem[\protect\citeauthoryear{Jose}{1965}]{1965AJ.....70..193J} Jose, P.~D.,\ 1965, \aj, {\bf 70}, 193
\bibitem[\protect\citeauthoryear{Klioner \& Kopeikin}{1992}]{klikop} Klioner, S.A. \& Kopeikin, S.M.,\  1992, \aj, {\bf 104}, 897
\bibitem[\protect\citeauthoryear{Klioner}{2003a}]{kl} Klioner, S.~A.,  2003a, {\it Astron. Astrophys}, {\bf 404}, 783
\bibitem[\protect\citeauthoryear{Klioner}{2003b}]{grem} Klioner, S.~A., 2003b, \aj, {\bf 125}, 1580
\bibitem[\protect\citeauthoryear{Kopeikin \& Sch\"afer}{1999}]{ks} Kopeikin, S.~M. \& Sch\"afer, G., 1999, \prd, {\bf
60}, id. 124002
\bibitem[\protect\citeauthoryear{Kopeikin}{2001}]{kop2001} Kopeikin, S.~M., 2001, {\it Astrophys. J. Lett.} {\bf 556}, L1
\bibitem[\protect\citeauthoryear{Kopeikin \& Mashhoon}{2002}]{km} Kopeikin, S.~M. \& Mashhoon, B., 2002, \prd, {\bf
65}, id. 064025
\bibitem[\protect\citeauthoryear{Kopeikin}{2004}]{kop-cqg} Kopeikin, S.~M.,  2004, {\it Class. Quantum Grav.} {\bf 21},
3251
\bibitem[\protect\citeauthoryear{Kopeikin \& Fomalont}{2006}]{kffp} Kopeikin, S.~M. \& Fomalont, E.~B.,\ 2006,
{\it Found. Phys.}, {\bf 36}, 1244
\bibitem[\protect\citeauthoryear{Kopeikin et al.}{ 2007}]{kpsv} Kopeikin, S.~M.,
Polnarev, A.~G., Sch{\"a}fer, G.
\& Vlasov, I.~Y.,\ 2007, {\it Physics Lett. A}, {\bf 367}, 276
\bibitem[\protect\citeauthoryear{Kopeikin \& Fomalont}{ 2007}]{kop-fom} Kopeikin, S.~M. \& Fomalont, E.~B.,\ 2007, {\it Gen. Rel. Grav.}, {\bf 39}, 1583
\bibitem[\protect\citeauthoryear{Kopeikin \& Makarov}{2007}]{kmak}Kopeikin, S.~M. \& Makarov, V.~V.,\ 2007, \prd, {\bf 75}, 062002
\bibitem[\protect\citeauthoryear{Kopeikin}{2009}]{kop09} Kopeikin, S.~M., \ 2009, {\it Physics Lett. A}, in press (arXiv:0809.3433)
\bibitem[\protect\citeauthoryear{Kosteleck\'y \& Potting}{1995}]{kos1}Kosteleck\'y, V.A. \& Potting, R., \ 1995, \prd, {\bf  51}, 3923
\bibitem[\protect\citeauthoryear{Landau \& Lifshitz}{1971}]{ll} Landau, L.~D. \& Lifshitz, E.~M., 1971, {\it The Classical Theory
of Fields} (Pergamon, Oxford)
\bibitem[\protect\citeauthoryear{Milani et al.}{2002}]{2002PhRvD..66h2001M} Milani, A.,
Vokrouhlick{\'y}, D., Villani, D., Bonanno, C.
\& Rossi, A.,\ 2002, \prd, {\bf 66}, 082001
\bibitem[\protect\citeauthoryear{Misner, Thorne \& Wheeler}{1973}]{mtw} Misner, C.~W., Thorne, K.~S. \& Wheeler, J.~A., 1973, {\it
Gravitation} (Freeman: New York)
\bibitem[\protect\citeauthoryear{Moyer}{2003}]{moy} Moyer, T.~D., 2003, {\it Formulation for Observed and
Computed Values of Deep Space Network Data Types for Navigation} (John Wiley
\& Sons: Hoboken)
\bibitem[\protect\citeauthoryear{Mukhanov}{2005}]{mukhanov} Mukhanov, V.~F.,\ 2005, {\it Physical
Foundations of Cosmology} (Cambridge University Press: Cambridge)
\bibitem[\protect\citeauthoryear{Ni}{2007}]{2007NuPhS.166..153N} Ni, W.-T.,\ 2007, {\it Nuclear Phys. B,
Proc. Suppl.}, {\bf 166}, 153
\bibitem[\protect\citeauthoryear{Pitjeva}{2008}]{pitjeva} Pitjeva, E.~V., 2008, private communication
\bibitem[\protect\citeauthoryear{Plowman \& Hellings}{2006}]{2006CQGra..23..309P} Plowman, J.~E. \& Hellings, R.~W.,\
2006, {\it Class. Quantum Grav.}, {\bf 23}, 309
\bibitem[\protect\citeauthoryear{Richter \& Matzner}{1982}]{1982PhRvD..26.2549R} Richter, G.~W. \& Matzner, R.~A.,\
1982, \prd, {\bf 26}, 2549
\bibitem[\protect\citeauthoryear{Richter \& Matzner}{1983}]{1983PhRvD..28.3007R} Richter, G.~W. \& Matzner, R.~A.,\
1983, \prd, {\bf 28}, 3007
\bibitem[\protect\citeauthoryear{Shapiro}{1964}]{shap} Shapiro, I.~I., 1964, {\it Phys. Rev. Lett.} {\bf 13}, 789
\bibitem[\protect\citeauthoryear{Standish \& Williams}{2006}]{standish} Standish, E.~M. \& Williams, J.~G., 2006, {\it Orbital ephemerides of the Sun, Moon, and Planets}, in: Explanatory Supplement to the Astronomical Almanac, Chapter 5, ed. P. K. Seidelmann, (U.S. Naval Observatory, Washington D.C.)
\bibitem[\protect\citeauthoryear{Teyssandier \& Le
Poncin-Lafitte}{2008}]{2008arXiv0803.0277T} Teyssandier, P. \& Le Poncin-Lafitte, C.,\ 2008, {\it Class. Quantum Grav.}, {\bf 25}, 145020
\bibitem[\protect\citeauthoryear{Turyshev, Shao \& Nordtvedt}{2004}]{2004CQGra..21.2773T} Turyshev, S.~G., Shao,
M. \& Nordtvedt, K.,\ 2004, {\it Class. Quantum Grav.}, {\bf 21}, 2773
\bibitem[\protect\citeauthoryear{Weisberg \& Taylor}{2005}]{wt}Weisberg, J.~M. \& Taylor, J.~H.,\ 2005, {\it The Relativistic
Binary Pulsar B1913+16: Thirty Years of Observations and Analysis}, in:
Binary Radio Pulsars, Eds. F.~A. Rasio and I.~H. Stairs, (San Francisco:
Astronomical Society of the Pacific), ASP Conference Series, {\bf 328}, pp.
25--31
\bibitem[\protect\citeauthoryear{Will}{1993}]{w-book} Will, C.~M., 1993, {\it Theory and Experiment in Gravitational
Physics} (Cambridge University Press: Cambridge)
\bibitem[\protect\citeauthoryear{Will}{2001}]{will-livrev}Will, C.~M., 2001, {\it Liv. Rev. Rel.}, 4, 4;
http://www.livingreviews.org/lrr-2001-4  (cited on Apr 7, 2008)
\bibitem[\protect\citeauthoryear{Williams, Turyshev \& Boggs}{2004}]{wtb}Williams, J.~G.,
Turyshev, S.~G., \& Boggs, D.~H.,\ 2004, {\it Phys. Rev. Lett.}, {\bf 93}, 261101
\end{thebibliography}
\end{document}